\documentclass[preprint,aps]{revtex4}
\usepackage{graphicx}

\begin{document}

\title{Epidemic spreading in dynamic small world networks}
\author{ Sheng Li\footnote{lisheng@sjtu.edu.cn}, Meng Meng\footnote{meng88\_2001@sjtu.edu.cn},
HongRu Ma\footnote{hrma@sjtu.edu.cn} }
\affiliation{Institute of Theoretical Physics, Shanghai JiaoTong University}
\date{November 8, 2004}
\begin{abstract}
In this paper, we use a series of small world networks to simulate the
epidemic spreading in the real world. To make our model more similar to the
real world, we employ a parameter $p_{move}$ to denote its moving
probability, which corresponds with the general mobility of individuals in
realistic social network. The networks keep the same small world
properties when they are varying. And the $SIRS$ model is used to simulate the
disease propagation. From these simulations we see the moving probability of
a small world network has a profound effect on the velocity for the spread
of an infection. The quantitative results show that the characteristic time $%
\tau $ of infection at initial stage decreases exponentially with $p_{move}$.
\end{abstract}

\maketitle

\section{Introduction}

Lots of works focus on the dynamics of an infectious disease
depending on the structure of population. There are two kinds of
classical ways to analyze this phenomenon in theoretics: Consider the
structure of population to be well mixed
populations\cite{Anderson, Murray0, Bailey, Landa} or spatially extended
subpolulations\cite{Mikhailov, Murray, Kallen, Fuentes}.
But both of them do not match the real populations very
well. The real world is neither well mixed nor lattices. Recently, large
amounts of statistics have suggested that the ideal model of the social
network should be "small world" model. Introduced by Watts and Strogatz\cite%
{Watts}, small world networks attempt to translate, into abstract model, the
complex topology of social interactions. Small worlds may play an important
role in the study of the influence of the network structure upon the
dynamics of many social processes such as disease spreading, biological
metabolizing, etc\cite{Albert}. However, in most of these works, only the
static small world networks were considered, which means the connections in
the network keep unchanged during the process\cite{Moore, Abramson}. The social processes are
dynamic and the social interactions are changing with time. Every individual
has a tendency to change its original contact with its acquaintances and
establish new relationship with others every day. Therefore, the study of
dynamic network becomes essential.

Using a newly-introduced parameter "moving probability" to depict
this tendency, we simulated the spread of epidemic disease in
small world networks under different moving conditions, and found
that the moving ability of a network has a great influence on the
spreading speed.

\section{Epidemic model}

Small world network model was put forward by Watts and Strogatz in 1998. To
adjust a single parameter $p$ from $0$ to $1$, we can get gradually changing
networks ranging from a regular lattice to a completely random graph. In this
model the parameter $p$ characterizes the degree of disorder of the network.
Its algorithm is as follows: Firstly, start with a ring lattice with $N$
nodes in which every node is connected to its first $K$ neighbors ($K/2$ on
each side). Secondly, every link connecting a node to a neighbor is then
randomly rewired to any other vertex of the system with the probability $p$.
However, self connections and duplicate edges are excluded. By varying $p$
from $0$ to $1$, we can see the network evolute from a regular network to a
completely random network progressively.

We make these small world networks motive through the following
steps: Firstly, construct a new small world network with the same
probability $p$ as a template. Then use a moving probability
$p_{move}$ to decide whether a connection between two nodes should
be reconnected or not. If a pair of nodes should be reconnected,
let them reconnect as what they do in the template networks. For
some node pairs have the same connection status in both original
network and template network, they will keep the same status after
moving. Therefore the "real" moving probability of our random
network model is $p\ast p_{move}$. The new network we get is
different from the original one but it keeps small-world
properties.

The interactions between the elements of the population are
described by a small world network. The links represent the
contact between subjects, and infection can only diffuse through
them. According to the realistic spreading of most kinds of
epidemic diseases (such as SARS), they usually have three
stages: susceptible($S$), infected($I$), and
refractory($R$). This is what the system, usually called as
$SIRS$, essentially means: the health condition of each individual
is described by a single dynamical variable adopting one of these
three values. Through contagion by an infected one, susceptible
elements can pass to the infected state. And after an infection
time $\tau _{I}$, infected elements pass to the refractory state,
which means that the infected ones are detected and then isolated
in quarantine. Refractory elements return to the susceptible state
after a recovery time $\tau _{R}$. We call it as SIRS for the
cycle
that a single element goes over. The contagion is possible only during the $%
S$ phase, and only by an $I$ element. During the $R$ phase, the elements
are immune and do not infect.

So each element is characterized by a time counter $\tau
_{i}(t)=0,1,...,\tau _{I}+\tau _{R}\equiv \tau _{0}$, describing its phase
in the cycle of the disease\cite{Abramson}. The epidemiological state $\pi _{i}(t)$ of the
element ($S$, $I$ or $R$) depends on this phase in the following way:%
\begin{eqnarray}
\pi _{i}(t) &=&S\hspace{1cm}if\ \tau _{i}(t)=0  \nonumber \\
\pi _{i}(t) &=&I\hspace{1cm}if\ \tau _{i}(t)\in [1,\tau _{I}] \\
\pi _{i}(t) &=&R\hspace{1cm}if\ \tau _{i}(t)\in [\tau _{I}+1,\tau _{0}]
\nonumber
\end{eqnarray}%

The state of an element in the next step depends on its current phase in the
cycle, and the states of its neighbors in the network. The rules of
evolution are the following:%
\begin{eqnarray}
\tau _{i}(t+1) &=&0\hspace{2cm}if\;\tau _{i}(t)=0\;and\;no\;infection\;occurs
\nonumber \\
\tau _{i}(t+1) &=&1\hspace{2cm}if\;\tau _{i}(t)=0\;and\;i\;becomes\;infected
\nonumber \\
\tau _{i}(t+1) &=&\tau _{i}(t)+1\hspace{0.7cm}if\;1\leq \tau _{i}(t)<\tau
_{0} \\
\tau _{i}(t+1) &=&0\hspace{2cm}if\;\tau _{i}(t)=\tau _{R}  \nonumber
\end{eqnarray}

With the fast development of modern vehicles, people are inclined to
changing their links with the original neighbors and establishing new
contact with others. Accordingly, another parameter $p_{move}$ is introduced
here to characterize the ``moving inclination" of individuals.
The networks move, i.e. change their original structures, with time processing,

Using the small world network model characterized by the SIRS mechanism, we
simulated the propagation of the epidemic disease in the population with
different $p_{move}$.

\section{NUMERICAL RESULTS}

We performed the numerical simulations on network with $N=10^{4}$ to $10^{6}$
vertices and with $K=20$ to $40$. What we noticed is the initial stage of
the infection. Normally, at the initial stage, the number of infected
elements grows in an exponential way.%
\begin{equation}
n_{inf}(t)\sim \exp (t/\tau )
\end{equation}%
where $\tau $ is the characteristic time of the growth. In the
simulations, we found the characteristic time decreases fast with
the increase of moving probability $p_{move}$. The results are
similar when disorder parameter $p$ varies in a proper range.

In order to get the characteristic time from numerical simulations
creditably, the time of the exponential growth should be long
enough. For
this purpose, in our results given below we set the disorder parameter as $%
p=0.01$, the nodes number as $N=10^{6}$ and $K=40$. The initial infected fraction $%
p_{inf}(0)$ is $0.000005$.

\begin{figure}[t]
\begin{center}
\includegraphics[width=0.43\linewidth]{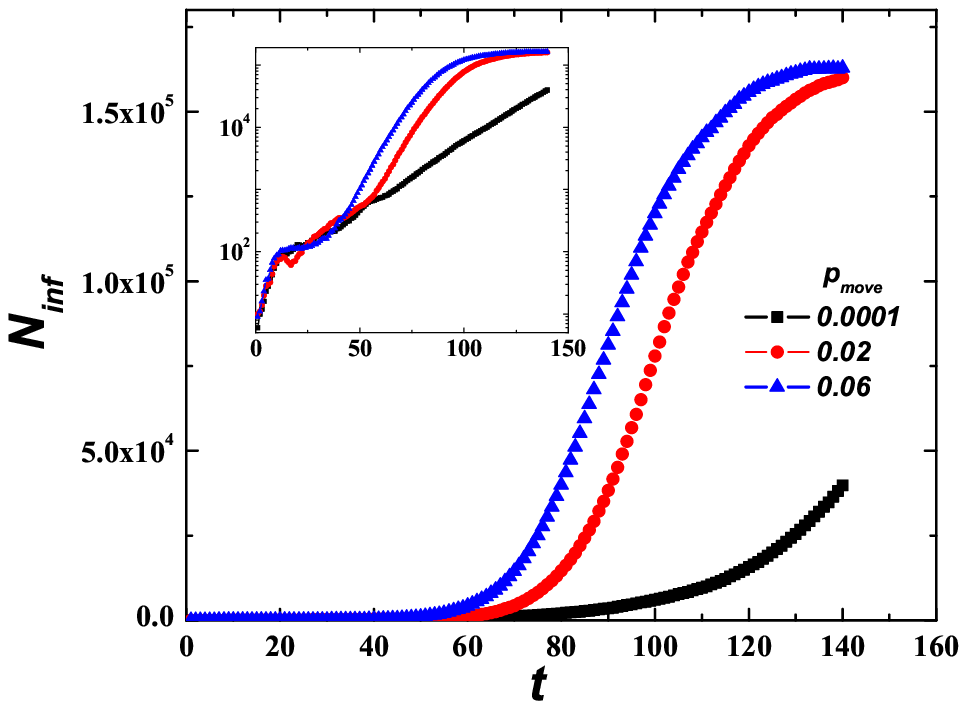}
\includegraphics[width=0.45\linewidth]{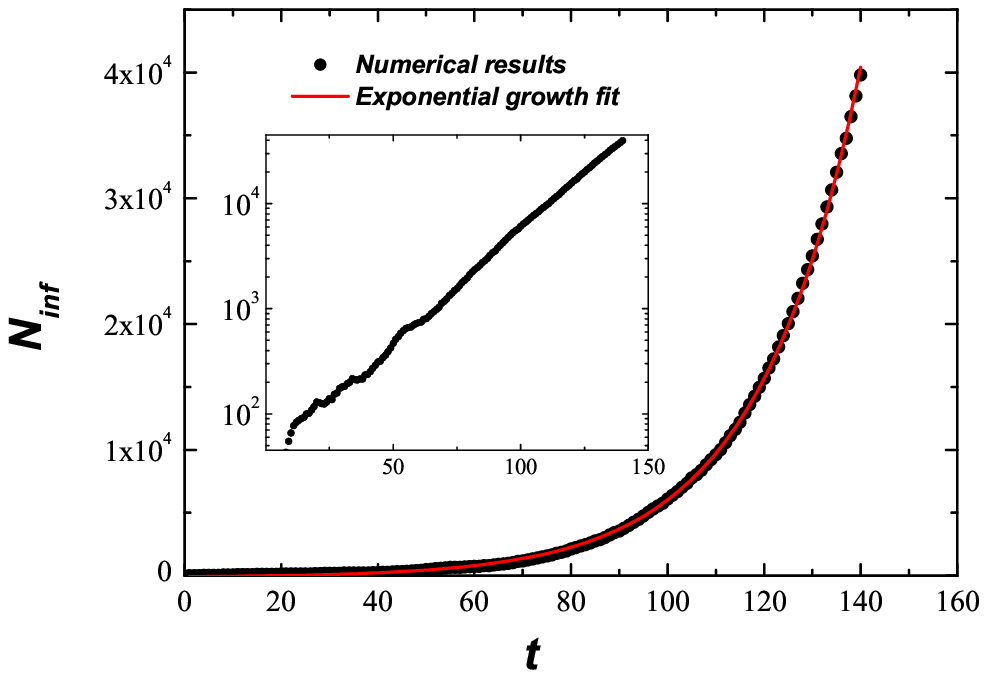}
\end{center}
\caption{Number of infected ones as a function of time. On the left three
series are shown, corresponding to different values of $p_{move}$. On the
right, we see the number of infected elements is an approximate exponential
function of time: $n_{inf}(t)\sim\exp(t/\protect\tau)$. And the small-scale
graph shows the relation between $\log(n_{inf}(t))$ and t. Here $p=0.0001,\, %
0.02$ and $0.06$. Other parameters are: $N=10^{6}$, $K=40$, $\protect%
\tau_{I}=6$, $\protect\tau_{R}=10$, $p_{inf}(0)= 0.000005$. }
\label{g1}
\end{figure}

We show in Fig.1 three time series displaying the number of infected
elements in the system as a function of time. The curves correspond to
systems with different values of the moving parameter: $p_{move}=0.0001$, %
$0.02$ and $0.06$. They have infection cycles with $\tau _{I}=6$
and $\tau _{R}=10$ and the result of $140$ days are shown. We can
see clearly that after a short period of time the number of
infected elements begins an approximately exponential growth with
time increasing.

At $p_{move}=0.0001$, where the network is relatively reluctant to
change its structure, the plot accords with the exponential fit
exactly and keeps increasing till up to the last day. However, at
$p_{move}=0.02$, where the network it represents becomes more
``active" and changeable, it grows more
rapidly with time and then, after the peak, enters a plateau. Similarly, at $%
p_{move}=0.06$, which means the most active network, it grows fast
with time first and then becomes constant at the day even earlier than the
former one does.

We also give out the relationship between $\log (n_{inf}(t))$ and $t$, and
then we see clearly the dependence of the growth of the infected people on
time. In the left small-scale graphs in Fig.1, we compare the growth
characters of the three different $p_{move}$. In the first several days when
infection begins, the three curves have almost the same slope. That is
because initially the number of infected elements is very small.
At this time the disease is
not widely distributed, which is closely related to the mutual contact
between individuals. Thus the three have almost the same growth rate with
time despite of their different $p_{move}$. After this period, the three
curves demonstrate different growth slopes. This is because the larger $%
p_{move}$ a network has, the quicker rate it has to change its structure,
which leads to the more rapid spreading of the disease. At last, the upper
two curves become ``flat", which mean that the epidemic spreading approaches
saturated state. In the right small-scale one, wiping off the first several
days and the plateau stage, the curve of $p_{move}=0.0001$ fits linear growth very well. Then
\begin{equation}
n_{inf}(t)\sim \exp (t/\tau ) %
\end{equation}
In the following
analysis, we take only the exponential growing stages into
account.

\begin{figure}[t]
\begin{center}
{\includegraphics[width=0.8\linewidth]{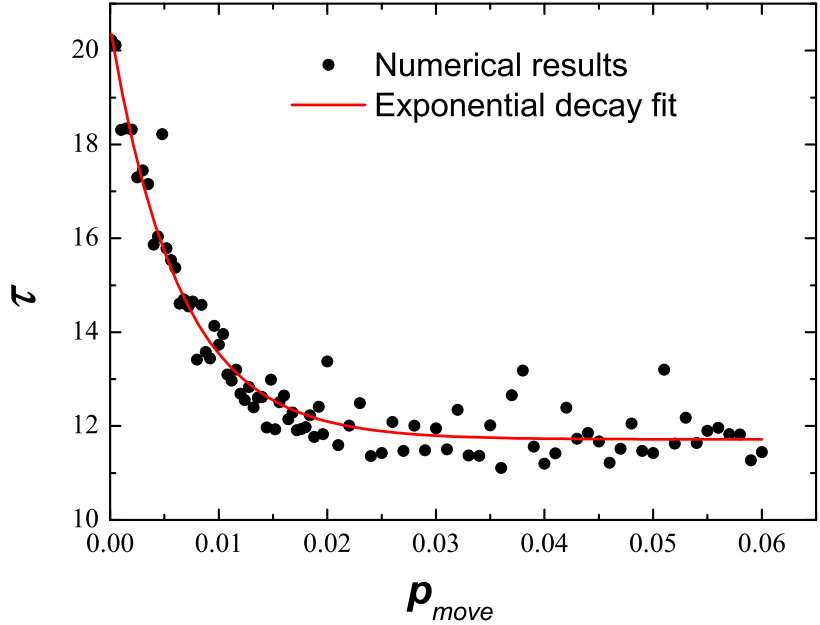}}
\end{center}
\caption{$\protect\tau$ as a function of $p_{move}$: $\protect\tau%
\sim\exp(-p_{move}/p_{0})$. Here $p_{0}$ is a constant. Each point corresponds to the
average of $20$ realizations. The line is of exponential decay fit. The formula is
$\tau =\tau _{0}+\tau _{1}e^{-\frac{p_{move}}{p_{mc}}}$, where
$\tau_{0}=11.72$, $\tau_{1}=8.77$ and $p_{mc}=0.0064$ is the characteristic moving
probability. Other parameters are $K=40$, $\tau_{I}=4$, $\tau_{R}=9$, $p=0.01$ and
$N=10^6$.
}
\label{g2}
\end{figure}

One can see that in the $\log$ scale figure, in the exponential growth stage,
the characteristic times of $p_{move}=0.06$ and of $p_{move}=0.02$ are almost the same,
but much smaller than
the characteristic time of $p_{move}=0.0001$. It means when the moving probability $p_{move}$
is large enough, the characteristic time of infection will converge.

From these results we see the epidemic spreading depends closely
on the moving probability of a social network. To analyze their
relationship more clearly, we extract in different $p_{move}$ the
time spans during which the plot growth accords with time
exponentially and record their characteristic time $\tau $. Then
we find that the characteristic time grow as an exponential decay
function of their respective moving probabilities in Fig 2.
\begin{equation}
\tau =\tau _{0}+\tau _{1}e^{-\frac{p_{move}}{p_{mc}}}
\end{equation}%
where $\tau _{0}$ is the characteristic time when the moving
probability is very larger; $p_{mc}$ is a characteristic moving
probability and $\tau _{0}+\tau _{1}$ is the characteristic time
when the network is static. Therefore, the spreading speed of the
infection increases exponentially when the network begins to move.
One should be noted that the characteristic moving probability
$p_{mc}$ ($0.0064$ with parameters given above) is very small,
which means that once a little freedom given to the network to
vary its connection, the infection speed increases markedly. What
is the inner mechanism in it that leads to this kind of functional
relationship between them and to what extent does this mechanism
play its role? We tried to explore this problem and analyzed these
simulated results.

\begin{figure}[t]
\begin{center}
{\includegraphics[width=0.8\linewidth]{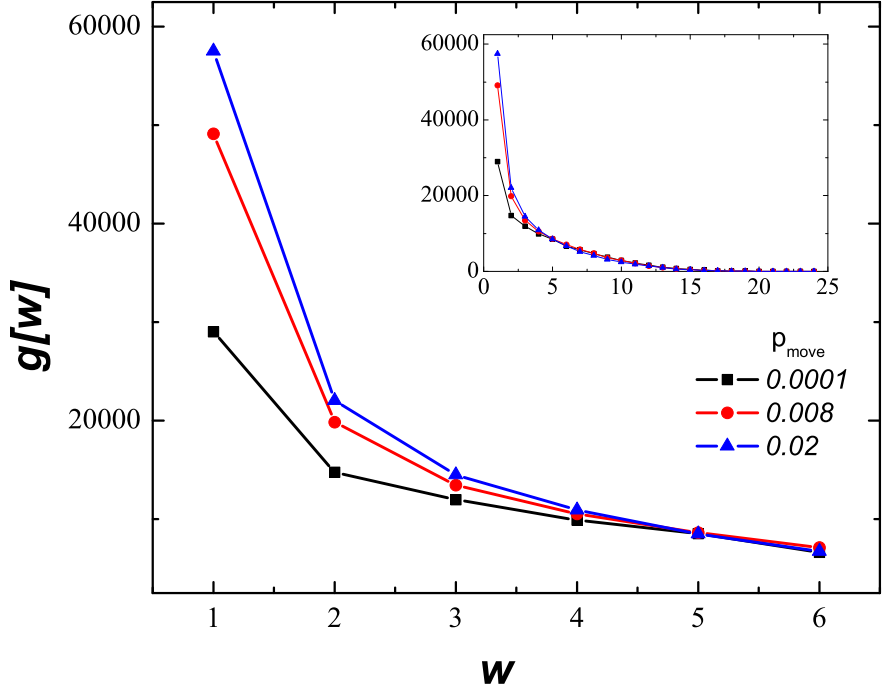}}
\end{center}
\caption{$g[w$] as a function of $w$; three curves under different $p_{move}$
are given. $g[w]$ is the number of healthy ones who contact with $w$
infected ones on the day when they have the same number of infected people.
Here the curve $p1$ is the one with $p_{move}$=0.0001, $p2$ with $p_{move}$%
=0.008, and $p3$ with $p_{move}=0.02$. We only show the curves from $w$=1 to
6. In the small-scale graph, we show the $g[w]$ with respect to $w$ from 1
to the largest possible one.}
\label{g3}
\end{figure}

First of all, we sought out those days pertaining to different $p_{move}$
models on which there are nearly the same number of infected people, say,
from 17000 to 18000, or, if there happens to be no such a number falling
into this span, from 16000 to 19000. Then we used an array $g[w]$ to record
the number of healthy ones who contact with w infected ones that day (here w
varies from 1 to 45). Analytically, we can give the expression of $n_{inf}$
of a given day as%
\begin{equation}
n_{inf}(t+1)=n_{inf}(t)+\sum_{w}[1-(1-p)^{w}]g[w]-\Delta
n_{ref}(t+1)
\end{equation}%
The last term $\Delta n_{ref}(t+1)$ refers to the number of
infected people that change to the refractory state that day.

In Fig.3, given the same $w$ ($w$ from 1 to 6), $g[w]$ varies
greatly according to different $p_{move}$: the greater the value
of $p_{move}$ is, the more $g[w]$ it has. That is the main reason
why epidemic disease spreads faster at a larger $p_{move}$. But from
the small-scale graph, we see that of different $p_{move}$
 $g[w]$ are in a
close range when $w$ becomes larger. This has little effect on the
result because when $w$ is large enough, the possibility of being
infected is very close to $1$. (The possibility of being infected
equals to $1-(1-p)^{w}$). So it is the part of small $w$ that
really counts.

\section{DISCUSSION}

Why does it work as an exponential function of $p_{move}$?
Unfortunately we can't figure out the underlying reason
analytically. As mentioned above, the result in Fig.3 can prove it
reasonable, but far from shedding light on the nature of the
phenomenon.

However, the result has its instructional meanings in reality in
some a way. When the epidemic disease emerges, what should we do to
effectively check its spread? We may expect to minimize its harm
by restricting the freedom of people's daily activities, that is,
alarm people not to go out and touch people as few as possible.
Then how about its effect? If our model works, the result in Fig.2
tells us that unless we can limit people's activities to an
extremely low degree ($p_{move}\approx 0$), it won't live up to
our expectations in controlling the disease spread. On the other hand,
in fig. 2, we see that the fluctuations of the characteristic moving
probability are not small. It also makes it hard to minimize the
spreading speed of infection by restricting the freedom of people's
activities.

In summary, we have found that the moving probability of a small
world network has a profound influence on the speed for the
spreading of an infection. The dynamic behavior of an SIRS
epidemic model is closely related to $p_{move}$. The
characteristic time of infection spreading in the initial stage
decreases with the increasing of moving probability $p_{move}$ in
an exponential way. This may be explained in some a sense by the
simulated results in FIG.3, where an effect of the social contact
change is observed.

\begin{acknowledgments}
This paper was supported by the National Science Foundation of
China under Grant No. 10105007, No. 10334020 and No. 90103035. .

\end{acknowledgments}

\end{document}